\UseRawInputEncoding
\documentclass[10pt,journal]{IEEEtran}
\IEEEoverridecommandlockouts
\usepackage{cite}
\usepackage{amsmath,amsfonts,amsthm,bm,amssymb} %
\usepackage[table,xcdraw]{xcolor}
\usepackage{algorithm}
\usepackage{algpseudocode} 
\usepackage{graphicx}
\usepackage{graphics}
\usepackage{epsfig}
\usepackage{footnote}
\usepackage{textcomp}
\PassOptionsToPackage{no-math}{fontspec}
\usepackage{xcolor}
\usepackage{tikz}  
\usepackage{makecell}
\usepackage{bm}
\usetikzlibrary{calc} 
\usetikzlibrary{arrows,shapes,chains} 
\usetikzlibrary{positioning, shapes.geometric}
\usepackage{multirow} %multirow for format of table
\usepackage{verbatim}
\usepackage{enumitem}
\def\BibTeX{{\rm B\kern-.05em{\sc i\kern-.025em b}\kern-.08em
		T\kern-.1667em\lower.7ex\hbox{E}\kern-.125emX}}
	
\usepackage{verbatim}
\usepackage{pgfplots}
\usepackage{hyperref}
\pgfplotsset{width=10cm,compat=1.9}
\pgfplotsset{
	every axis legend/.style={
		cells={anchor=center},
		inner xsep=3pt,inner ysep=2pt,
		nodes={inner sep=2pt,text depth=0.1em},
		anchor=north east,
		shape=rectangle,
		fill=white,draw=green,
		font=\footnotesize%\tiny
	},
}
\usepackage{verbatim}
\usepackage{booktabs}
\usepackage{subcaption}
\usepackage{fancyhdr}
\usepackage{svg}
\normalsize
\usepgfplotslibrary{external}
\tikzexternaldisable  % Explicitly disable
	
\graphicspath{{./figs/}}
\begin{document}
\title{Quasi-BP for BCH Codes and its Optimization\\
%	\footnotesize 
%	\thanks{applicable funding agency here. If none, delete this.}
}
\begin{comment}
\author{\IEEEauthorblockN{Guangwen Li}
	\IEEEauthorblockA{\textit{College of Information \& Electronics} \\
		\textit{Shandong Technology and Business University}\\
		Yantai, China \\
		lgwa@sdu.edu.cn}
}
\end{comment}

\author{Guangwen Li
\thanks{G.Li is with School of Information \& Electronics, Shandong Technology and Business University, Yantai, China e-mail: lgw.frank@sdtbu.edu.cn}% <-this % stops a space
}
\maketitle
\begin{abstract}
This paper proposes a quasi-belief propagation decoder for BCH codes that systematically integrates domain knowledge--specifically, channel noise variance, the cyclic property of the codes, and the deliberate redundancy in their parity-check matrices--to enable efficient iterative decoding. We rigorously formalize this parallelizable decoder within an information-theoretic framework by tracking mutual information evolution through the constituent variable and check decoders, thereby validating the use of scattered EXIT charts as a tool for optimizing the decoder's parameters. At each iteration, an input dilation operation expands the set of messages, while a subsequent merging operation accelerates mutual information growth, ensuring rapid convergence. The proposed decoder achieves decoding performance approaching that of LDPC codes with comparable rate and blocklength, effectively pioneering the feasible deployment of BP-like decoding for high-density parity-check codes. The generality and robustness of the scheme are demonstrated through extensive simulations across codes of varying rates and blocklengths.
\end{abstract}


%\IEEEpeerreviewmaketitle
\section{Introduction}
Since Shannon's groundbreaking work in information theory \cite{shannon1948mathematical}, channel coding has remained a cornerstone of modern reliable telecommunications. Among the broad family of linear block codes, low-density parity-check (LDPC) codes are distinguished by their exceptional error-correcting capability and their potential to asymptotically approach the Shannon limit under canonical belief propagation (BP) decoding \cite{gallager62, mackay96}. In complexity-sensitive applications, BP decoding is often supplanted by min-sum (MS) variants~\cite{zhao05, jiang06} such as normalized min-sum (NMS)~\cite{chen2005reduced}. However, these BP-like methods are suboptimal for finite-length LDPC codes due to short cycles in their Tanner graphs, and the error-floor~\cite{richardson2003error} in the high signal-to-noise ratio (SNR) region imposes further constraints on the design of the parity-check matrix $\mathbf{H}$. 

In comparison, classical linear block codes with high-density parity-check (HDPC) matrices---such as Bose-Chaudhuri-Hocquenghem (BCH) and Reed-Solomon (RS) codes---are widely used in storage and deep-space communications due to their large minimum distances and effective hard-decision decoding schemes. Soft BP decoding performs poorly for BCH codes because preserving their algebraic structure introduces numerous short cycles in the corresponding Tanner graph. Therefore, a key objective in reviving BCH codes for next-generation communication systems or Internet of Things (IoT) networks \cite{zhang2024improved} is to devise a new decoding scheme that achieves performance comparable to LDPC BP decoding in terms of frame error rate (FER), computational complexity, throughput, and latency.

On the other hand, ordered statistics decoding (OSD) \cite{Fossorier1995} was proposed to narrow the gap between BP and maximum-likelihood (ML) decoding for short codes. Recent acceleration strategies for classical BCH codes and short LDPC codes \cite{Yue2021} aim to increase throughput by triggering early stopping criteria or reducing the number of test error patterns through thresholding. Despite these efforts, the inherently sequential nature of OSD and its variants limits their suitability as primary decoders in throughput- or delay-sensitive scenarios.

The rapid advancement of deep learning has inspired new perspectives in error-correction coding. Nachmani et al. \cite{nachmani16} pioneered the unrolling of iterative BP decoding into a neural network (NN), giving rise to neural belief propagation (NBP) by assigning trainable weights to the edges of the Tanner graph. Other NN architectures, such as convolutional neural networks (CNNs) and recurrent neural networks (RNNs), have also been explored for various linear codes \cite{nachmani18, liang18}. A two-stage decimation process was introduced in a recent study \cite{buchberger21} where the likely bits are first identified using an NN and then followed by list decoding with NBP. However, such customized operation undermines the parallell processing capacity of BP decoding.

The analysis and design of iterative decoders were revolutionized by the introduction of extrinsic information transfer (EXIT) charts by ten Brink~\cite{ten1999convergence}. EXIT charts provide a powerful semi-analytical tool to visualize decoding convergence behavior asymptotically by tracking the mutual information (MI) exchange between soft-input soft-output modules~\cite{ten2002convergence}. Building upon this foundation, the concept of scattered EXIT (S-EXIT) charts~\cite{ebada2018scattered} was introduced to address the limitations of conventional EXIT analysis in the finite-length regime, and was subsequently formalized as a general optimization tool for finite-length LDPC codes.

In prior work~\cite{li2025effective}, we demonstrated that NMS can effectively decode BCH codes by exploiting the redundancy in $\mathbf{H}$ and the cyclic property of the codes. The present work extends this approach to a canonical BP-like framework to probe the fundamental limits of decoding capability when channel noise variance is known. Our main contributions are as follows:

\begin{itemize}
    \item A parallelizable quasi-BP decoder for BCH codes is proposed, achieving FER performance comparable to canonical BP decoding for LDPC codes.
    \item The decoder's architecture is analyzed using MI concepts, with key parameters optimized via S-EXIT charts, clearly exposing the internal progression of iterative decoding.
    \item Extensive simulations validate the feasibility of the proposed quasi-BP decoder for high-rate longer BCH codes.
\end{itemize}

The remainder of this paper is organized as follows. Section~\ref{preliminary} introduces preliminaries on BP variants, EXIT, and S-EXIT charts. Section~\ref{sec:motivations} details the decoding method and its optimization. Section~\ref{simulations} presents FER results for various BCH codes, along with a brief complexity analysis and discussion. Finally, Section~\ref{conclusions} concludes the paper.
\section{Preliminaries}
\label{preliminary}

Consider a binary message row vector $\mathbf{m} = [m_i]_{i=1}^K$ encoded into a codeword $\mathbf{c} = [c_i]_{i=1}^N$ via $\mathbf{c} = \mathbf{m}\mathbf{G}$ over GF(2), where $K$ and $N$ denote the message and codeword lengths, respectively, and $\mathbf{G}$ is a full-rank generator matrix.

Each codeword bit $c_i$ is mapped to an antipodal symbol using binary phase shift keying (BPSK) modulation: $s_i = 1 - 2c_i$. After transmission over an additive white Gaussian noise (AWGN) channel with noise $n_i \sim \mathcal{N}(0,\sigma^2)$, the received soft sequence is
\begin{equation}
	y_i = s_i + n_i, \qquad i=1,\dots,N,
\end{equation}
which is fed into a dedicated decoder to estimate the original codeword. The log-likelihood ratio (LLR) for the $i$-th bit is thus
\begin{equation}
	l_{i,{\text{ch}}} = \log\frac{p(y_i\mid c_i=0)}{p(y_i\mid c_i=1)} = \frac{2y_i}{\sigma^2}.
	\label{eq:llr}
\end{equation}
 Owing to the preserved symmetry property of BP decoders, the all-zero codeword is assumed hereafter in simulations for simpler notation without loss of generality.

\subsection{BP Variants and Neural Counterparts}

A code is defined by a bipartite Tanner graph underlying its parity-check matrix $\mathbf{H}$. The graph contains $N$ variable nodes and $M \geq N-K$ check nodes if redundancy is allowed, with an edge connecting variable node $j$ to check node $i$ whenever $H_{ij} \neq 0$.

Assuming a flooding schedule with at most $l_{\max}$ iterations, canonical BP exchanges LLR messages along the Tanner graph edges. The variable-to-check message at iteration $t$ ($t = 1,\dots,l_{\max}$) is
\begin{equation}
	l_{\nu_i \to c_j}^{(t)} = l_{i,{\text{ch}}} + \sum_{\substack{ p \in \mathcal{C}(i)\setminus j}} l_{c_p \to \nu_i}^{(t-1)},
	\label{eq:v2c}
\end{equation}
where $\mathcal{C}(i)\setminus j$ denotes the neighbors of $\nu_i$ except $c_j$, and $l_{c_p \to \nu_i}^{(0)} = 0$. The check-to-variable message flowing in the opposite direction is
\begin{equation}
	l_{c_j \to \nu_i}^{(t)} = 2\tanh^{-1}\!\Biggl(
	\prod_{\substack{ q \in \mathcal{V}(j)\setminus i}}
	\tanh\!\Bigl(\frac{l_{\nu_q \to c_j}^{(t)}}{2}\Bigr)
	\Biggr),
	\label{eq:c2v}
\end{equation}
with $\mathcal{V}(j)\setminus i$ being the neighbors of $c_j$ except $\nu_i$.

After each iteration, the a-posteriori LLR of the $i$-th bit is
\begin{equation}
	l_{\nu_i}^{(t)} = l_{i,{\text{ch}}} + \sum_{\substack{ p \in \mathcal{C}(i)}} l_{c_p \to \nu_i}^{(t-1)},
	\label{eq:aposteriori}
\end{equation}
which yields the tentative hard decision
\begin{equation}
	\widehat{c}_i^{(t)} = 
	\begin{cases}
		0, & \operatorname{sgn}\!\bigl(l_{\nu_i}^{(t)}\bigr)=+1,\\[2pt]
		1, & \text{otherwise}.
	\end{cases}
	\label{eq:hard_decision}
\end{equation}
Decoding terminates early to reduce complexity if $\widehat{\mathbf{c}}^{(t)}\mathbf{H}^{\mathsf T}=\mathbf{0}$, where $\widehat{\mathbf{c}}^{(t)}=[\widehat{c}_i^{(t)}]_{i=1}^N$. Otherwise, set $t \leftarrow t + 1$ and execute another iteration through \eqref{eq:v2c}--\eqref{eq:hard_decision} until the $l_{\max}$-th iteration is reached.

To avoid the costly $\tanh$ and $\tanh^{-1}$ operations, the min-sum approximation replaces \eqref{eq:c2v} with
\begin{equation}
	l_{c_j \to \nu_i}^{(t)} = s_{c_j \to \nu_i}^{(t)}\; \phi_{c_j \to \nu_i}^{(t)},
	\label{eq:ms}
\end{equation}
where
\begin{align}
	s_{c_j \to \nu_i}^{(t)} &= \prod_{\substack{ q \in \mathcal{V}(j)\setminus i}}
	\operatorname{sgn}\!\bigl(l_{\nu_q \to c_j}^{(t)}\bigr),\label{eq:sign}\\
	\phi_{c_j \to \nu_i}^{(t)} &= \min_{\substack{ q \in \mathcal{V}(j)\setminus i}}
	\bigl|l_{\nu_q \to c_j}^{(t)}\bigr|. \label{eq:min}
\end{align}
Since the approximation in \eqref{eq:ms} consistently overestimates the exact value in \eqref{eq:c2v}, NMS or offset min-sum (OMS) decoders apply a multiplicative weight or an additive offset, respectively, to suppress the overestimation effect.

Furthermore, when trainable weights are assigned to or shared among the edges of a specific network trellis of BP variants, the resulting decoder is termed NBP.

\subsection{EXIT and S-EXIT Charts}

The core principle of EXIT charts is to decompose a complex iterative decoding process into two component decoders---for LDPC codes, the variable node decoder and the check node decoder. The behavior of iterative decoding can then be predicted by tracking the evolution of the average MI exchanged between these two components. Practically, at iteration $t$, the average MI is defined as the expectation of MI values,
\begin{equation}
	\bar I^{(t)}
	\triangleq
	\mathbb{E}_{e,r}
	\!\left[
	I\!\left( L_{e,r}^{(t)} \right)
	\right]
	=
	1
	-
	\mathbb{E}_{e,r}
	\left[
	\log_2
	\left( 1 + e^{-L_{e,r}^{(t)}} \right)
	\right],
	\label{eq:def_MI}
\end{equation}
with the set of individual MI samples denoted as
\begin{equation}
	\mathcal{I}^{(t)}
	=
	\Big\{
	I\!\left( L_{e,r}^{(t)} \right)
	\;\Big|\;
	e \in \mathcal{E},\;
	r \in \mathcal{R}
	\Big\},
\end{equation}
where $\mathcal{E}$ denotes the set of edges in the Tanner graph,
$\mathcal{R}$ the set of channel realizations, and $L_{e,r}^{(t)}$ the LLR message
passed along edge $e$ at iteration $t$ for the $r$-th channel realization. A transfer characteristic curve relating the average a-priori MI to the average extrinsic MI can be plotted for each component. The two curves form a `tunnel' on the chart; one can intuitively predict successful convergence if the tunnel remains open, allowing iterations to advance smoothly toward a high-MI point, which implies a low error rate.

For finite-length codes, the averaging operation obscures the intrinsic randomness of the decoding process. S-EXIT chart analysis therefore retains the individual MI samples before averaging, yielding a cloud of points in the MI plane that captures the distribution of convergence behavior.
\section{Method and Optimization}
\label{sec:motivations}

\subsection{Quasi-BP Iterative Decoding}

The proposed quasi-BP decoder mitigates detrimental message correlation from dense short cycles by exploiting the redundancy in $\mathbf{H}$ and the code's cyclic property. Automorphisms enable input dilation, and additional rows are incorporated into $\mathbf{H}$ to facilitate iterative decoding while preserving the code's algebraic structure. Building on canonical BP and enhanced NMS~\cite{li2025effective}, the decoding procedure comprises the following steps:

\begin{itemize}
	\item \textbf{Preprocessing:} Optimize the targeted $\mathbf{H}$ by determining a redundancy factor $\delta_1$, which measures the desired level of redundancy. Additionally, specify a dilation factor $\delta_2$ that controls the extent of input dilation.
	
	\item \textbf{Initialization:} Initialize the LLR message $L_{\nu_i}^{(0)}$ using \eqref{eq:llr}, and set all check-to-variable messages $m_{c_j \rightarrow \nu_i}^{(0)}$ to zero.
	
	\item \textbf{Variable Node Update:} This update consists of two phases: message alignment and merging (Phase One), followed by input dilation (Phase Two).
	
	\textit{Phase One -- Alignment and Merging:} Split and align the incoming messages $m_{c_j \rightarrow \nu_i}^{(t-1)}$ by applying a block-wise cyclic reversion $\phi^{-1}(\cdot)$, which restores them to a size compatible with the pre-dilation input. This yields $\delta_2$ message blocks $\phi(m^{(t-1,w)})$. The pre-dilation input is then updated according to
	\begin{equation}
		L_{\nu_i}^{(t)} = L_{\nu_i}^{(t-1)} + \beta \sum_{w = 1}^{\delta_2} \sum_{j \in \mathcal{C}(i)} \phi^{-1}\!\big(m^{(t-1,w)}\big)_{c_j \rightarrow \nu_i},
		\label{quasi_bp_variable_node_update}
	\end{equation}
	where $\beta$ is a merging weight to be optimized.
	
	\textit{Phase Two -- Input Dilation:} Leverage the automorphism property of the BCH code to dilate the updated input $L_{\nu_i}^{(t)}$ by a factor $\delta_2$, and dispatch the dilated messages to the check nodes. For each variable node $\nu_i$ in the dilated representation,
	\begin{equation}
		m_{\nu_i \rightarrow c_j}^{(t)} = \phi\big(L_{\nu_i}^{(t)}\big), \quad j \in \mathcal{C}(i).
	\end{equation}
	
	\item \textbf{Check Node Update:} Compute $m_{c_j \rightarrow \nu_i}^{(t)}$ identically to the canonical BP update in \eqref{eq:c2v}.
	
	\item \textbf{Hard Decision and Termination:} Form a tentative hard decision based on $L_{\nu_i}^{(t)}$ from \eqref{quasi_bp_variable_node_update} using \eqref{eq:hard_decision}. If $\widehat{\mathbf{c}}^{(t)}\mathbf{H}^{\mathsf T} = \mathbf{0}$, terminate early; otherwise, proceed to the next iteration.
\end{itemize}

For a chosen $l_{\max}$, the sole parameter $\beta$ in quasi-BP can be efficiently optimized by maximizing the output mutual information $I_{_{E,C}}$ (denoted later) after the final iteration via a bisection line search.

\subsection{EXIT View of Quasi-BP}
\begin{figure}[htbp]
	\centering
	\begin{subfigure}[t]{0.48\columnwidth}
		\centering
		\includegraphics[width=\linewidth]{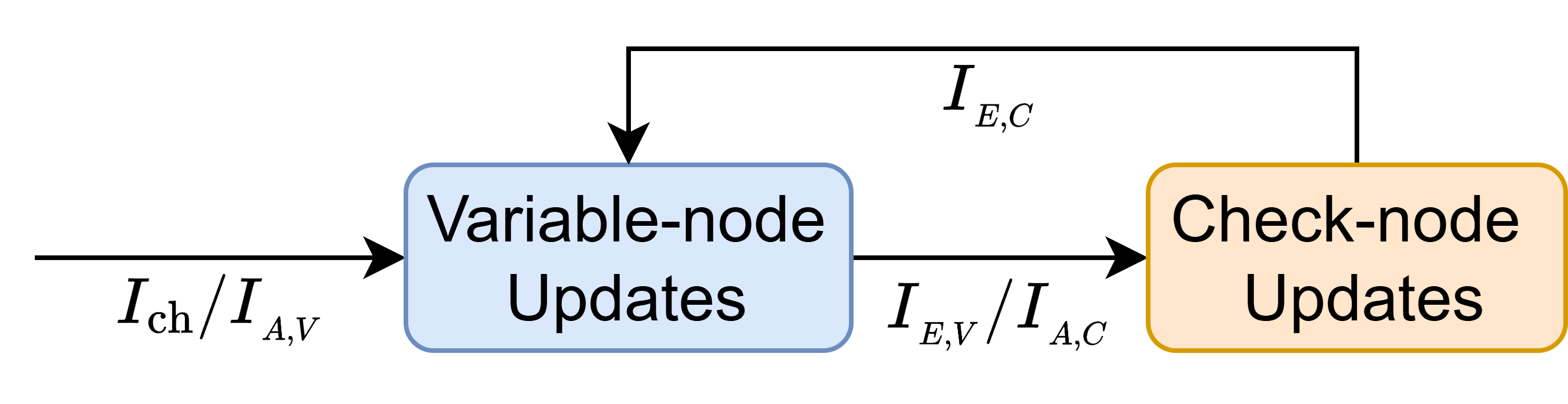}
		\caption{Canonical BP}
		\label{fig:bp_sub1}
	\end{subfigure}
	\hfill
	\begin{subfigure}[t]{0.48\columnwidth}
		\centering
		\includegraphics[width=\linewidth]{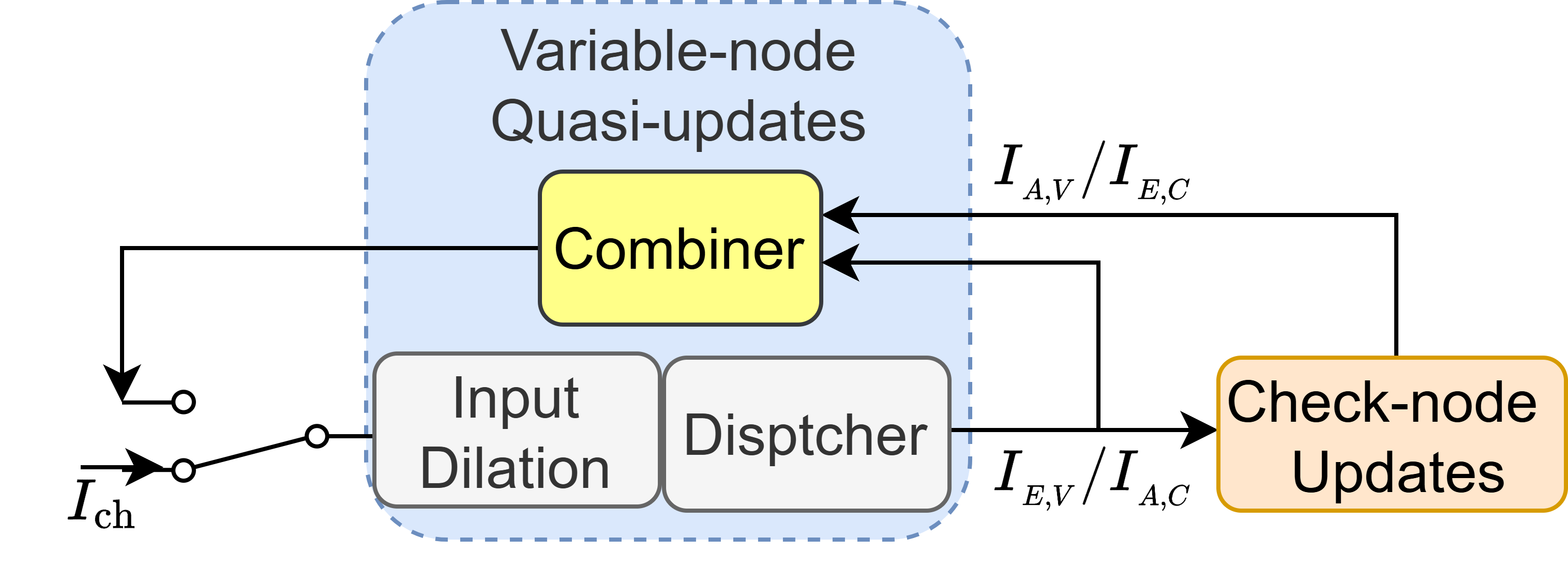}
		\caption{Quasi-BP}
		\label{fig:quasi_bp_sub2}
	\end{subfigure}
	\caption{Comparison of BP and quasi-BP in EXIT view.}
	\label{fig:information_view_both}
\end{figure}

Compared with canonical BP in  Fig.~\ref{fig:bp_sub1}, the operations at variable node side of quasi-BP  in Fig.~\ref{fig:quasi_bp_sub2} include message alignment and merging (labeled as `Combiner'), input dilation, and message dispatching. Among these three components, input dilation and message dispatching preserve MI evaluation; only the message merging operation within the combiner substantially increases MI. The input switch in Fig.~\ref{fig:quasi_bp_sub2} indicates that the channel input $I_{\text{ch}}$ is employed only once during initialization, with $I_{_{E,V}}^{(0)} = I_{\text{ch}}$. Thereafter, the switch toggles, $I_{\text{ch}}$ no longer contributes, and we have $I_{_{E,V}}^{(t+1)} = T_{_V}(I_{_{E,V}}^{(t)}, I_{_{A,V}}^{(t)})$. The check node update enforces $I_{_{E,C}}^{(t)} = T_{_C}(I_{_{A,C}}^{(t)})$.

In general, both mappings $T_{_V}(\cdot)$ and $T_{_C}(\cdot)$ are analytically intractable for finite-length codes, although the consistency relations $I_{_{A,C}}^{(t)} = I_{_{E,V}}^{(t)}$ and $I_{_{A,V}}^{(t)} = I_{_{E,C}}^{(t)}$ hold, where $I_{_{A,V}}$ ($I_{_{E,V}}$) and $I_{_{A,C}}$ ($I_{_{E,C}}$) denote the input (output) MI metrics of the variable nodes and check nodes, respectively. Consequently, we resort to S-EXIT charts analysis to approximate the metrics $I_{_{E,V}}$ and $I_{_{E,C}}$. Specifically, we define three types of expectation approximations as follows:
\begin{align}
	\mathbb{E}_e[\cdot ] &\approx \frac{1}{|\mathcal{E}|}\sum_{e\in \mathcal{E}}\log_2\!\big(1 + e^{-L_{e,r}^{(t)}}\big), \label{collapse_edges} \\
	\mathbb{E}_r[\cdot ] &\approx \frac{1}{|R|}\sum_{r\in R}\log_2\!\big(1 + e^{-L_{e,r}^{(t)}}\big), \label{collapse_realization} \\
	\mathbb{E}_{e,r}[\cdot ] &\approx \frac{1}{|\mathcal{E}||R|}\sum_{e\in \mathcal{E},r\in R}\log_2\!\big(1 + e^{-L_{e,r}^{(t)}}\big). \label{collapse_edges_realization}
\end{align}

At the $t$-th iteration, \eqref{collapse_edges} and \eqref{collapse_realization} generate S-EXIT clouds by collapsing over edges and channel realizations, respectively, while \eqref{collapse_edges_realization} yields a mean value by collapsing over both dimensions. The evolution curves of $I_{_{E,V}}$ and $I_{_{E,C}}$ are thus obtained by substituting \eqref{collapse_edges_realization} into \eqref{eq:def_MI} to approximate the LLR messages exchanged between variable and check nodes across iterations at each SNR, which are then leveraged to optimize key parameters of quasi-BP, as detailed in the following.
\subsubsection{Selection of \(l_{\max}\)}

\begin{figure}[htbp]
	\centering
	\begin{subfigure}{0.485\linewidth}
		\resizebox{\linewidth}{!}{\includegraphics[width=\linewidth]{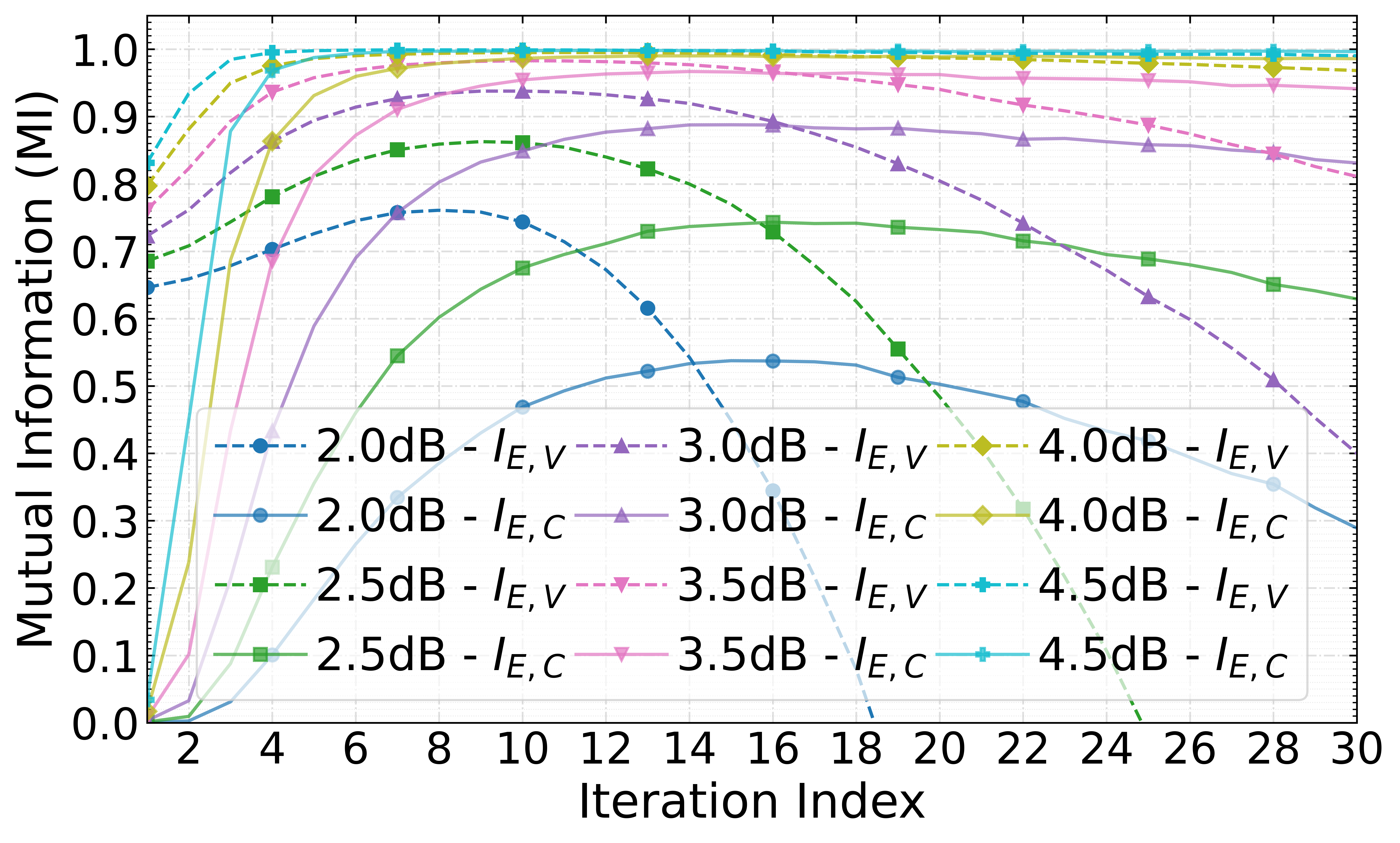}}
		\caption{Evolution of $I_{_{E,V}}$ and $I_{_{E,C}}$}
		\label{fig:optimize_max_iteration}
	\end{subfigure}
	\hfill
	\begin{subfigure}{0.5\linewidth}
		\resizebox{\linewidth}{!}{ \includegraphics[width=\linewidth]{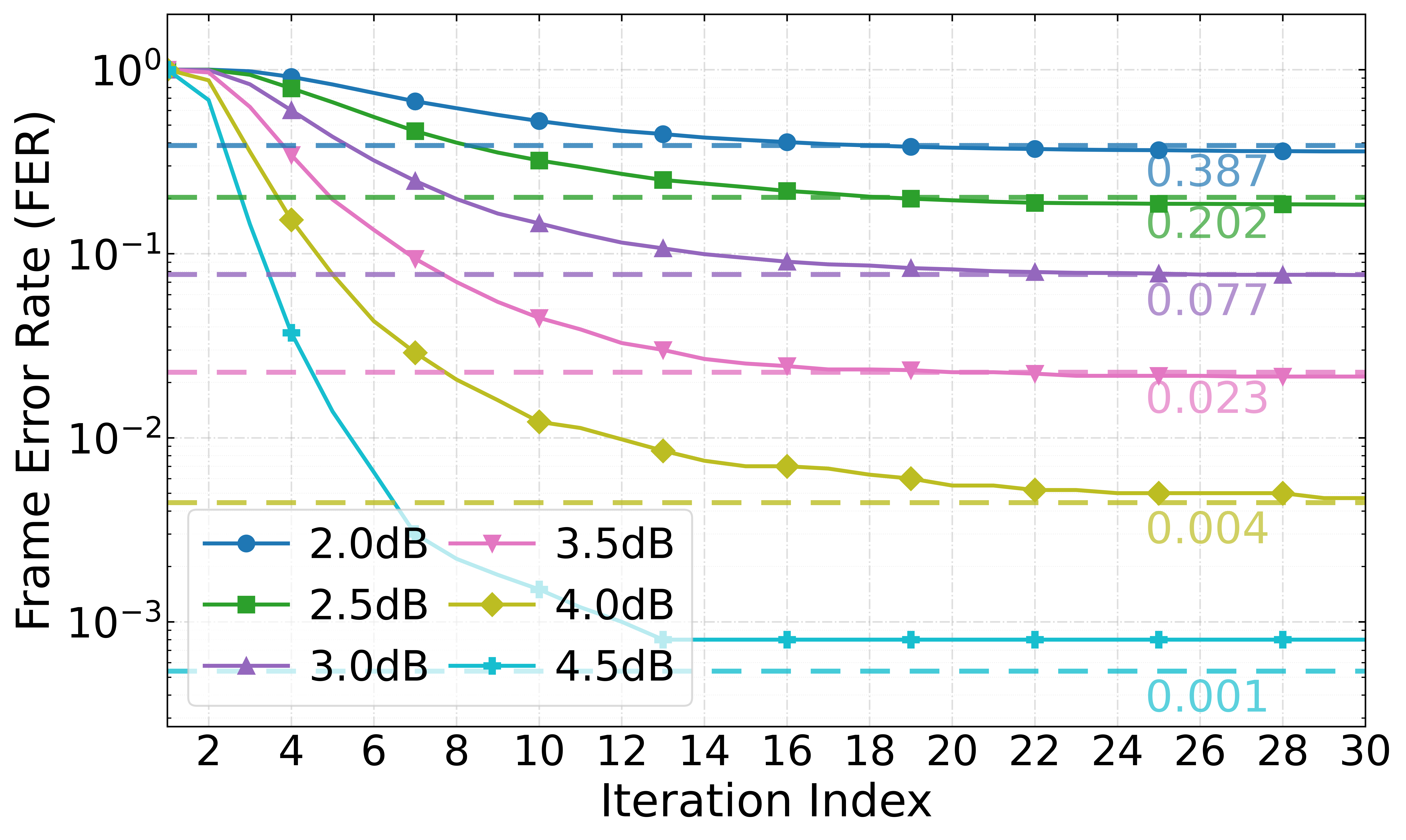}}
		\caption{FER estimated via $\mathbb{E}_e[\cdot ]<\tau$}
		\label{fig:MI_FER}
	\end{subfigure}
	\caption{Association of $I_{_{E,V}}$ and $I_{_{E,C}}$ with $l_{\max}$ optimization and FER estimation under ($l_{\max}$, $\delta_1$, $\delta_2$) = (30, 2, 15) for BCH(127,64) code.}
	\label{fig:optimize_max_iteration_full}
\end{figure}

As shown in Fig.~\ref{fig:optimize_max_iteration}, $I_{{E,V}}$ and $I{_{E,C}}$ climb, saturate, and decline asynchronously, diverging after their peaks. Per \eqref{collapse_edges_realization}, the decline stems from a few decoding failures whose penalties dominate the MI approximation. it may be misinterpreted as performance deterioration with more iterations.
With early termination, FER improves monotonically with $l_{\max}$, but practical applications must balance complexity and delay. We therefore establish a criterion for selecting $l_{\max}$: take the SNR where FER $\approx 0.1$ under sufficient iterations (e.g., 3.0 dB for BCH(127,64) code in Fig.~\ref{fig:optimize_max_iteration}), and set $l_{\max}$ to the intersection index of $I_{{E,V}}$ and $I{{E,C}}$ at that SNR plus two iterations (yielding $l{\max}=18$).

Compared to $I_{_{E,V}}$, the $I_{_{E,C}}$ metric is more informative as it quantifies the extent to which parity-check equations are satisfied—an aspect that lies at the very core of linear block code decoding. Thus, it will underscore our following discussions.

\subsubsection{Connection Between MI and FER}

Select an appropriate threshold $\tau$ applicable to all SNRs, such that the proportion of received sequence realizations satisfying $\mathbb{E}_e[\cdot ] < \tau$ can accurately approximate the FER conventionally secured by collecting at least 100 decoding errors per SNR point.  Fig.~\ref{fig:MI_FER} clearly shows that the estimated FERs obtained by counting the cases where the edge-collapsed MI falls below threshold $\tau = 0.7$ gradually approach the dashed curves achieved conventionally per SNR point. Further S-EXIT simulations verify that as the iteration index increases, a typical bimodal phenomenon emerges: the $\mathbb{E}_e[\cdot ]$ values of successfully decoded realizations gradually tend toward unity, while those of failed realizations degrade toward large negative values. Consequently, the selection of $\tau$ is quite lenient; the resulting proportion is so insensitive that $\tau = 0.5$ or $0.9$ yields almost identical FER estimates. Furthermore, the setting of $l_{\max} = 18$ is well justified after observing the marginal improvement in FER beyond that point.

Notably, more than $10^5$ codeword samples are required to stabilize the dashed curve at SNR = 4.5 dB in Fig.~\ref{fig:MI_FER}. In comparison, only 4,000 realizations are involved in drawing the solid curves in Fig.~\ref{fig:MI_FER}, at the cost of only a minor discrepancy. Hence, from the perspective of easing FER stabilization, counting cases where $\mathbb{E}_e[\cdot ] < \tau$ is more favorable for reducing simulation variation than the conventional method of directly counting frame errors.
\subsubsection{Selection of $\delta_1$ and $\delta_2$}
\begin{figure}[htbp]
	\centering
	\begin{subfigure}{0.48\linewidth}
		\resizebox{\linewidth}{!}{\includegraphics[width=\linewidth]{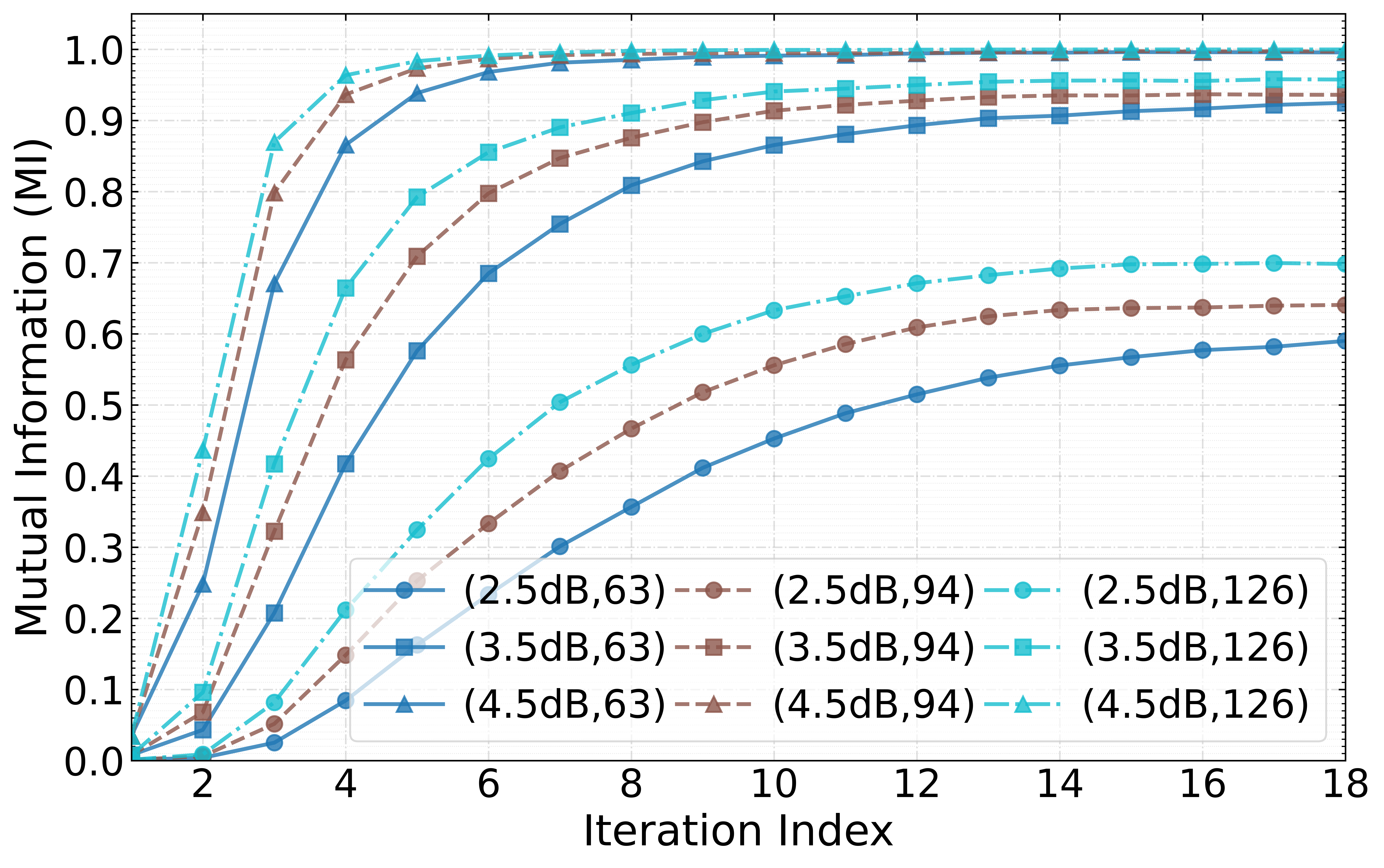}}
		\caption{$\delta_1$ varies, $\delta_2=9$}
		\label{fig:optimize_redundancy_factor}
	\end{subfigure}
	\hfill
	\begin{subfigure}{0.50\linewidth}
		\resizebox{\linewidth}{!}{ \includegraphics[width=\linewidth]{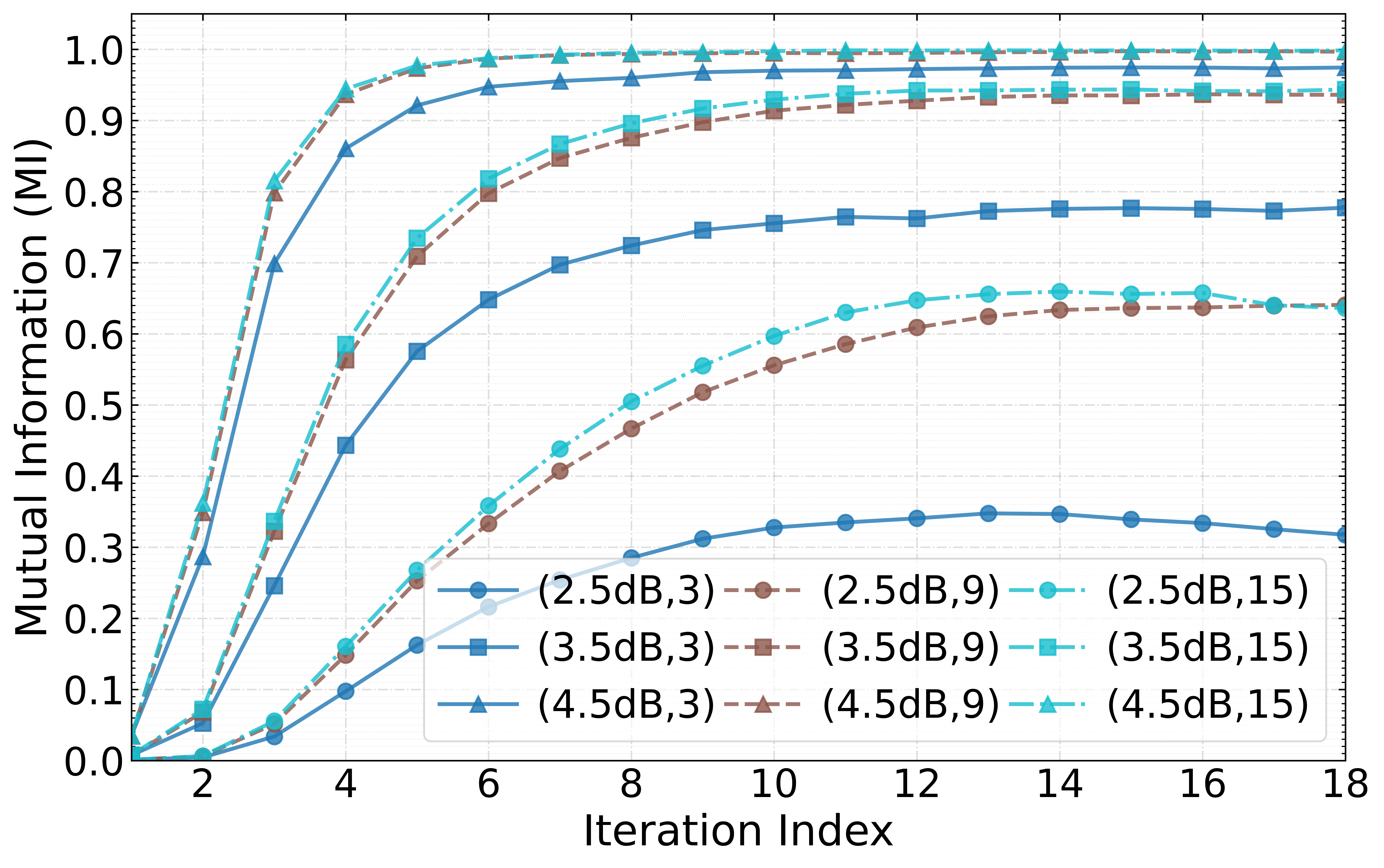}}
		\caption{$\delta_2$ varies, $\delta_1=1.5$.}
		\label{fig:optimize_input_dilation}
	\end{subfigure}
	\caption{Impact of $\delta_1$ and $\delta_2$ on $I_{E,C}$  with respect to iteration index  for BCH(127,64) code parameterized by varying SNR.}
	\label{fig:delta_impact}
\end{figure}

Fixing $\delta_2 = 9$, the evolution of $I_{E,C}$ with respect to iteration index is examined in Fig.~\ref{fig:optimize_redundancy_factor} for various redundancy levels. Higher SNR or a larger $\delta_1$ (equivalently, $63\delta_1$ rows in $\mathbf{H})$, yields faster MI growth as well as higher final values after the allotted iterations, although the resulting gain from large $\delta_1$ narrows at high SNR points. Similarly, Fig.~\ref{fig:optimize_input_dilation} demonstrates that, with $\delta_1 = 1.5$ fixed, larger $\delta_2$ values consistently achieve faster and higher MI across all SNR points. However, a saturation effect is observed: increasing $\delta_2$ from 9 to 15 yields only marginal MI improvement across the SNR region of interest, indicating that further increases beyond 15 offer diminishing returns, even if a wider range of $\delta_2$ values were available.
Larger $\delta_1$ or $\delta_2$ entails higher computational complexity, necessitating a trade-off between complexity and FER performance when selecting these parameters.

\section{Simulation and Analysis}
\label{simulations}
We evaluate FER/BER performance of parallelizable iterative decoders for BCH(127,64), BCH(127,99), BCH(255,239), and CCSDS LDPC(128,64) \cite{channelcodes} codes. Non-parallelizable decoders (e.g., those requiring sorting or $\mathbf{H}$ reduction) are excluded. Parameters $(\delta_1, \delta_2) = (2, 15)$ are fixed for all BCH codes, with $l_{\max}$ specified per scheme. Source code is available on GitHub.\footnote{\url{https://github.com/lgw-frank/Short\_BCH\_quasi-BP\_decoding}}
\subsection{Potentials of Quasi-BP}
The LDPC(128,64) code--matching the BCH(127,64) code in rate and blocklength--serves as a reference for FER comparison. Larger $l_{\max}$ yields negligible FER improvement for all decoders considered.

\begin{figure}[htbp]
	\centering
	\begin{tikzpicture}
	\begin{semilogyaxis}[
		width=1.1\textwidth, 
		height=0.7\textwidth,      
		scale = 0.3,
			% Font size settings
		label style={font=\footnotesize},      % For axis labels (xlabel, ylabel)
		tick label style={font=\scriptsize},      % For tick numbers 
		xlabel={$E_b/N_0$(dB)},
		ylabel={FER/BER},
		xmin=0.75, xmax=5.0,
		ymin=1e-5, ymax=1,
		xtick={1.0,1.5,2,2.5,3,3.5,4,4.5,5.0},
		ytick={1e-4,1e-3,1e-2,1e-1,1},
		legend pos = south west,
		ymajorgrids=true,
		xmajorgrids=true,
		minor x tick num=4,
		minor grid style={dotted, gray!50},
		grid style=dashed,
		legend style={font=\tiny\selectfont, fill opacity=0.7, text opacity=1},
		]
		% BP-RNN(5) - BER - BCH
		\addplot[
		color=green!70!black,
		mark=triangle*,
		mark size=1.5pt,
		dashed
		] coordinates {
			(1.00,1.3e-1)
			(2.00,1e-1)
			(3.00,7.5e-2)
			(4.0,4e-2)
			(5.0,1.3e-2)
			(6.0,3e-3)
		};
		\addlegendentry{BP-RNN(5) (BCH) BER\cite{nachmani18}}
		
		% NMS(8) - FER - BCH
		\addplot[
		color=orange,
		mark=diamond*,
		mark size=1.5pt,
		very thin
		] coordinates {
			(1.00,0.93495)
			(1.50,0.84881)
			(2.00,0.70094)
			(2.50,0.50150)
			(3.00,0.29223)
			(3.50,0.13072)
			(4.00,0.0428)
			(4.50,0.0098)
			(5.00,0.0014)
		};
		\addlegendentry{NMS(8) (BCH) \cite{li2025effective}}
		% Quasi-BP(20) - FER - BCH
		\addplot[
		color=magenta,
		mark=pentagon*,
		mark size=1.5pt,
		very thin
		] coordinates {
			(2.00,0.38692)
			(2.50,0.2024)
			(3.00,0.07708)
			(3.50,0.02269)
			(4.00,0.00444)
			(4.50,0.00054)
		};
		\addlegendentry{Quasi-BP(20) (BCH) FER}
		% NBP(50) - FER - LDPC
		\addplot[
		color=red,
		mark=square*,
		mark size=1.2pt,
		very thin
		] coordinates {
			(1.0,0.91)
			(1.5,0.71)
			(2.0,0.4)
			(2.5,0.2)
			(2.9,0.09)
			(3.0,0.07)
			(3.25,0.035)
			(3.65,0.01)
			(3.75,0.007)
			(4.0,0.003)
			(4.75,1e-4)
		};
		\addlegendentry{NBP(50) (LDPC) FER \cite{buchberger21}}
		
		% BP(40) - FER - LDPC
		\addplot[
		color=blue,
		mark=*,
		mark size=1.5pt,
		very thin
		] coordinates {
			(1.0,0.82)
			(1.5,0.6)
			(2.0,0.31)
			(2.5,0.15)
			(3.0,0.056)
			(3.5,0.013)
			(4.0,0.0021)
			(4.5,2e-4)
			(5.0,2e-5)
		};
		\addlegendentry{BP(40) (LDPC) FER \cite{channelcodes}}        
	\end{semilogyaxis}
\end{tikzpicture}   
	\caption{FER/BER of BCH(127,64) and LDPC(128,64) codes.}
	\label{fig_bch_ldpc_fer_ber}
\end{figure}
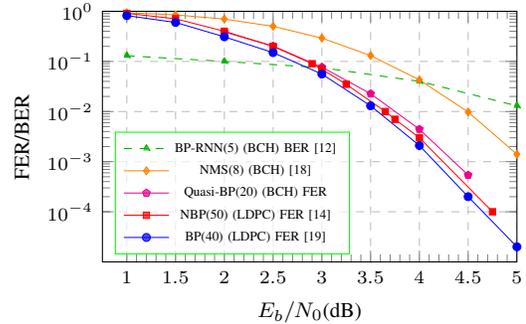
As shown in Fig.~\ref{fig_bch_ldpc_fer_ber}, for the BCH code, the dashed bit error rate (BER) curve of the neural BP-RNN decoder \cite{nachmani18} (FER data unavailable in the literature) exhibits a flat slope, indicating limited competitiveness due to an unadapted $\mathbf{H}$ and the absence of cyclic property exploitation. Although NMS with $l_{\max}=8$ \cite{li2025effective} achieves significantly better FER than BP-RNN, the proposed quasi-BP surpasses NMS by more than 0.6 dB at the cost of higher $l_{\max}$ and computational complexity, with the prerequisite of known noise variance. For the CCSDS LDPC(128,64) code, the NBP decoder with $l_{\max}=50$ leads the quasi-BP decoder for the BCH(127,64) code by only 0.1 dB, while traditional BP with $l_{\max}=40$ employing layered message scheduling (as opposed to flooding) provides an additional 0.1 dB gain.

It is observed the NBP and BP decoders for the LDPC code, along with the quasi-BP decoder for the BCH code, exhibit nearly identical slopes, indicating that the FER gap is not expected to widen with SNR. Furthermore, in the high-SNR region where LDPC codes suffer error floors, BCH codes are favored due to their excellent minimum distance and dense parity-check structure, which effectively eliminate such floors.

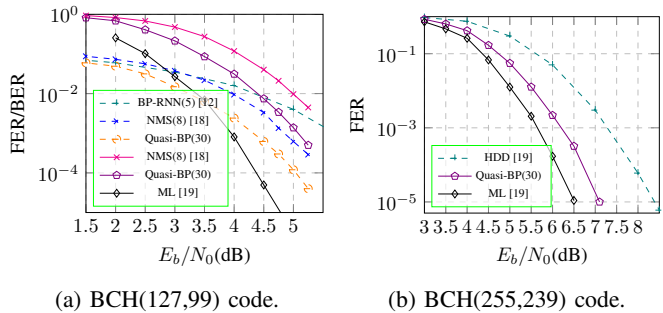
\begin{figure}[htbp]
	\centering
	\begin{subfigure}{0.49\linewidth}
		\resizebox{\linewidth}{!}{	\centering
\begin{tikzpicture}
	\begin{semilogyaxis}[
		scale = 0.5,
		xlabel={$E_b/N_0$(dB)},
		ylabel={FER/BER},
		xmin=1.5, xmax=5.5,
		ymin=1e-5, ymax=1.,
		xtick={1.0,1.5,2,2.5,...,5.0},
		legend pos = south west,
		ymajorgrids=true,
		xmajorgrids=true,
		grid style=dashed,
		legend style={font=\fontsize{6}{7}\selectfont,fill opacity = 0.7,text opacity = 1}
		]
		%BER 
		\addplot[
		color=teal,
		mark=+,
		dashed
		]
		coordinates {
			(1.0,0.08)
			(2.0,0.06)
			(3.0,0.035)
			(4.0,0.016)
			(5.0,0.004)
			(6.0,5.5e-4)
		};
		\addlegendentry{BP-RNN(5) \cite{nachmani18}}
		
		%0 plot for NMS (BER)
		\addplot[
		color=blue,
		mark=x,
		dashed,
		]
		coordinates {
			(1.00,0.10024)(1.50,0.08735)(2.00,0.07330) (2.50,0.05643) (3.00,0.03756) (3.50,0.02182) (4.00,0.00945) (4.50,0.00333)
			(4.75, 0.00135)
			(5.00, 0.00062)
			(5.25, 0.00029)
		};	
		\addlegendentry{NMS(8) \cite{li2025effective}}
		\addplot[
		color=orange,
		mark=halfcircle,
		dashed]
		coordinates {
			(1.00,0.07816)
			(1.50,0.05963)
			(2.00,0.04903)
			(2.50,0.03146)
			(3.00,0.01490)
			(3.50,0.00632)
			(4.00,0.00241)
			(4.50,0.00062)
			(4.75, 0.0003)
			(5.00,0.00012)
			(5.25,0.00004)
		};	
		\addlegendentry{Quasi-BP(30) }
		%0 plot for NMS (FER)
		\addplot[
		color=magenta,
		mark=x,
		very thin
		]
		coordinates {
			(1.00, 0.97777)
			(1.50, 0.93065)
			(2.00, 0.82760)
			(2.50, 0.68020)
			(3.00, 0.47643)
			(3.50, 0.27441)
			(4.00, 0.11829)
			(4.50, 0.04015)
			(4.75, 0.02102)
			(5.00, 0.00969)
			(5.25, 0.00447)
		};	
		\addlegendentry{NMS(8) \cite{li2025effective}}
		
		\addplot[
		color=violet,
		mark=pentagon,
		very thin
		]
		coordinates {
			(1.00,0.96667)
			(1.50,0.82000)
			(2.00,0.68667)
			(2.50,0.40800)
			(3.00,0.21400)
			(3.50,0.08583)
			(4.00,0.03118)
			(4.50,0.00745)
			(4.75,0.0034)
			(5.00, 0.00138)
			(5.25,0.0005)
		};	
		\addlegendentry{Quasi-BP(30) }
		
		%0.5plot for ML (FER)
		\addplot[
		color=black,
		mark=diamond,
		]
		coordinates {
			(2.00, 2.551e-01)
			(2.50, 1.020e-01)
			(3.00, 2.680e-02)
			(3.50, 6.907e-03)
			(4.00, 8.074e-04)
			(4.50, 4.958e-05)
			(5.00, 2.984e-06)
		};	
		\addlegendentry{ML \cite{channelcodes}}
	\end{semilogyaxis}
\end{tikzpicture}	}
		\caption{BCH(127,99) code.}
		\label{fig_127_99_bch_fer_ber}
	\end{subfigure}
	\hfill
	\begin{subfigure}{0.49\linewidth}
		\resizebox{\linewidth}{!}{	\centering
\begin{tikzpicture}
		\begin{semilogyaxis}[
			scale = 0.5,
			xlabel={$E_b/N_0$(dB)},
			ylabel={FER},
			xmin=3.0, xmax=8.5,
			ymin=5e-6, ymax=1.,
			xtick={3.0,3.5,4.0,4.5,5.0,...,8.0},
			legend pos = south west,
			ymajorgrids=true,
			xmajorgrids=true,
			grid style=dashed,
            legend style={font=\fontsize{6}{7}\selectfont,fill opacity = 0.7,text opacity = 1}
			]
%BER 
\addplot[
color=teal,
mark=+,
dashed
]
coordinates {
(3.0,0.95)
(4.0,0.75)
(5.0,0.3)
(6.0,0.05)
(7.0,3e-3)
(8.0,6e-5)
(8.5,6e-6)
};
\addlegendentry{HDD \cite{channelcodes}}
\addplot[
color=violet,
mark=pentagon,
very thin
]
coordinates {
(3.00,0.85000)
(3.50,0.63636)
(4.00,0.41875)
(4.50,0.16917)
(5.00,0.05537)
(5.50,0.01267)
(6.00,0.00218)
(6.50,3.2e-4)
(7.1,1e-5)
};	
\addlegendentry{Quasi-BP(30)}

%0.5plot for ML (FER)
\addplot[
color=black,
mark=diamond,
]
coordinates {
(3.00, 7.246e-01)
(3.50, 4.673e-01)
(4.00, 2.604e-01)
(4.50, 6.793e-02)
(5.00, 1.256e-02)
(5.50, 2.054e-03)
(6.00, 1.708e-04)
(6.50, 1.093e-05)
};	
\addlegendentry{ML \cite{channelcodes}}
		\end{semilogyaxis}
	\end{tikzpicture}	}
		\caption{BCH(255,239) code.}
		\label{fig_255_239_bch_fer}
	\end{subfigure}
	\caption{FER/BER for BCH(127,99) and BCH(255,239) codes.}
	\label{fig_127_99_255_239_bch_fer}
\end{figure}

For higher-rate BCH(127,99) code, the FER and BER performance of quasi-BP is shown in Fig.~\ref{fig_127_99_bch_fer_ber}.
In terms of BER (dashed lines), BP-RNN lags behind NMS by at least 0.6 dB at $\mathrm{BER} = 10^{-3}$, and the performance gap widens with increasing SNR due to the shallower slope of the former decoder. For both FER and BER, the proposed quasi-BP leads NMS by at least 0.5 dB. However, the FER of quasi-BP still lags more than 1 dB behind the ML performance.

For BCH(255,239) code, the FER performance of quasi-BP is shown in Fig.~\ref{fig_255_239_bch_fer}. Quasi-BP is only about 0.6 dB away from ML performance at $\mathrm{FER} = 10^{-4}$, and due to similar slopes, this gap is not expected to increase with SNR. In comparison, despite its simplicity, the hard-decision decoding (HDD) algorithm based on Berlekamp-Massey (BM) operates only in serial mode and lags behind quasi-BP by about 1 dB, with the gap widening at higher SNRs. Finally, the extensible architecture of quasi-BP enables it to decode high-rate longer BCH codes effectively without any modification.

Compared with canonical BP, even ignoring the dilation and merging operations, quasi-BP incurs a complexity multiplier of approximately $\delta_1 \delta_2$ per iteration, resulting in higher worst-case computational complexity unless $\delta_1$ and $\delta_2$ are carefully designed and constrained using the S-EXIT tool to minimize performance loss. Fortunately, the fast decoding convergence observed in most SNR regions (as seen in Fig.~\ref{fig:MI_FER}) guarantees a substantial reduction in the average number of iterations, thereby lowering both average computational complexity and average decoding delay.

\subsection{Reflections on Quasi-BP}

It is well known that NMS decoding \cite{chen2005reduced} for LDPC codes can approach canonical BP performance with only a marginal gap. However, NMS lags behind quasi-BP for BCH codes by more than 0.5 dB, as demonstrated in the previous subsection. We conjecture that the root cause lies in the density of $\mathbf{H}$: any row of the BCH parity-check matrix is significantly denser than that of LDPC codes. For BCH codes, the diversified values returned by a check node in quasi-BP--typically involving more than ten non-zero variable nodes--are sharply reduced to binary decisions in NMS. In contrast, for LDPC codes, the smaller number of non-zero variable nodes per check node results in a milder reduction, yielding less performance loss.

As shown in the simulations above, despite the superior performance of quasi-BP, a considerable gap remains to reach ML decoding. Consequently, concatenating quasi-BP with OSD, that is, triggering OSD only upon quasi-BP failure, can achieve ML  without significantly increasing latency.
\section{Conclusions}
\label{conclusions}

This paper demonstrates that quasi-BP  represents a significant step forward in realizing efficient parallel iterative soft decoding for BCH codes, achieving performance that approaches that of their closest rivals--LDPC codes. In doing so, it challenges the conventional belief that BP-like variants, long dominated by LDPC codes, can hardly be applied effectively to HDPC codes. Since each bit of the received sequence undergoes identical processing operations throughout quasi-BP decoding, the architecture exhibits a valuable property that facilitates straightforward implementation on modern hardware accelerators such as GPUs or TPUs. Furthermore, S-EXIT charts are employed to investigate parameter optimization in quasi-BP, enabling a favorable trade-off between decoding performance and computational complexity.
For long BCH codes, where more than 50 variable nodes are typically involved in each check node update, maintaining the balance between precision and numerical stability in the consecutive multiplication of $\tanh$ functions is worthy of future investigation.
\bibliography{main}
\end{document}